\documentclass[prd,superscriptaddress,nofootinbib,preprint,tightenlines,showpacs]{revtex4}
\usepackage{graphicx}
\usepackage{dcolumn}
\usepackage{bm}
\usepackage{amssymb}
\usepackage{latexsym}
\usepackage{mathrsfs}
\usepackage{amsmath}
\usepackage{epsfig}
\usepackage{subfigure}
\usepackage[dvips]{color}
\usepackage{hhline}

\newcommand{\be}{\begin{equation}}
\newcommand{\ee}{\end{equation}}
\newcommand{\bea}{\begin{eqnarray}}
\newcommand{\eea}{\end{eqnarray}}
\newcommand{\al}{\alpha}
\begin{document}
\title{Correlations between low energy leptonic CP violation and leptogenesis in the light of recent experiments.}
\author{H. Zeen Devi}
\email{zeen@imsc.res.in}
\affiliation{The Institute of Mathematical Sciences, Taramani, Chennai 600113, India}
\begin{abstract}
Leptogenesis is the most favourable mechanism for generating the observed baryon 
asymmetry of the Universe (BAU) which implies CP violation in the high energy scale. The low energy leptonic CP violation is expected to be observed in the neutrino oscillations and $0\nu 2 \beta$ decay  experiments. Generally it is not possible to connect both the CP violations. Here we revisit the issue of connecting the two in flavoured leptogenesis scenario within the Type I seesaw in the light of recent neutrino oscillation and {\it Planck} data. With the recent precise measurements of $\theta_{13}$ and BAU we are able to find new correlations  between the low and high energy CP violating phases when leptogenesis occurs at temperature  between $10^9$ to $10^{12}$ GeV and there is no contribution to CP violation from the heavy neutrino sector.   
\end{abstract}
\maketitle
\section{Introduction}
The recent {\it Planck} experiment \cite{Ade:2013zuv} gives baryon density($\Omega_{b}h^2$) to be $0.02205\pm0.00028$ in the $68\%$ confidence level and the present value of baryon to entropy ratio is estimated to be $Y_B=(8.294-8.508)\times10^{-11}$ from the relation $\eta=3.81\times10^{-9}\Omega_{b}h^2$. This baryon number asymmetry can be generated by the process of thermal leptogenesis \cite{Fukugita:1986hr}. It is in this context that the Type I seesaw model becomes very interesting. It can explain the smallness of the neutrino mass as well as satisfy the Sakharov conditions necessary for successful baryogenesis. In addition to the Standard Model(SM) particles, three right handed (RH) heavy neutrinos which are singlet under the SM gauge group are added to generate a light neutrino mass in Type I seesaw model. These singlet heavy neutrinos are of Majorana type and therefore can decay into a particle and as well as to an anti-particle. Such lepton number violating decays with different decay rates for the particles and for the anti-particles can give rise to a net CP asymmetry of the particular flavour of the final state lepton. These CP asymmetries can survive only when the decays are out-of-equilibrium. Therefore a CP violating out-of equilibrium decay of the heavy right-handed neutrinos can produces a lepton asymmetry, which is then converted into the baryon asymmetry by the $B+L$ violating sphaleron process \cite{Kuzmin:1985mm}. Leptogenesis occurs at a high energy scale of temperature $T\simeq M_1$, where $M_1$ is the mass of the heavy RH neutrino.

In low energy neutrino physics the CP violating parameters are expressed in terms of three phases in the neutrino mixing matrix, the Dirac phase $\delta$, and two Majorana phases $\alpha_1$ and $\alpha_2$. The Dirac phase will be measured in long baseline experiments and the recent very precise measurement of sizeable neutrino reactor angle $\theta_{13}$ by \cite{An:2012eh,Ahn:2012nd} has made it within the reach of these future experiments ( \cite{Feldman:2012qt} and references therein). The Majorana phases does not appear in the oscillation measurements  but can be observed in the effective Majorana mass in neutrinoless double beta ($0\nu2\beta$) decay experiments. For a detailed discussion on $0\nu2\beta$ decays see Ref.\cite{Rodejohann:2012xd}. 

CP violation in low energy sector may come from the phases appearing in the left handed  and the right handed fields, whereas the phases responsible for leptogenesis are those appearing in the right handed fields. So observation of low energy CP violation does not necessarily also mean high energy CP violation. But there has been many efforts in the past to connect them in the flavoured leptogenesis scenario \cite{Pascoli:2006ie,Branco:2006ce,Pascoli:2006ci}. 
Thermal leptogenesis is basically studied in a 'single flavour' regime where all the lepton flavours are considered to be indistinguishable and all the charged lepton Yukawa coupling are out-of-equilibrium, which is true only for temperatures $T\geq 10^{12}GeV$. However, if we consider CP to be an exact symmetry of the right handed (RH) sector then the total CP asymmetry, which is sum of the asymmetry in each flavour goes to zero. Therefore leptogenesis occurring at temperature greater than $10^{12}\,GeV$ cannot generate a non-zero CP asymmetry for this class of models. 
 But for temperatures within $10^{9}\leq T \leq 10^{12} GeV$ the charged $\tau$ Yukawa couplings come into equilibrium and the lepton asymmetry can still survive even in the presence of CP invariant RH sector. It was shown in \cite{Branco:2006ce,Pascoli:2006ci} that in these class of models with the CP invariant RH neutrino sector it is possible to connect the low to the high scale CP violation when flavour effects are considered.  
In \cite{Branco:2006ce} the authors studied the correlation between the low CP violating phases for the observed baryon asymmetry for different light neutrino mass hierarchies around the central values of the neutrino oscillation parameters considering the value of $\sin \theta_{13}$ to be 0.01 and 0.15. In \cite{Pascoli:2006ci} the authors drew a correlation between low energy CP invariant term
$J_{CP}$ and the baryon asymmetry of the universe.    
In the present work we have scanned for the range of recent $1\sigma$ values of the oscillation parameters \cite{GonzalezGarcia:2012sz} and carried the analysis for the three possible light neutrino mass spectrum and have observed a considerable change from the earlier results. This becomes interesting in the light of the recent measurements of precise value of $\theta_{13}$ and stringent limits on BAU by the {\it Planck} and WMAP 9. We would also like to mention here that we have considered the right handed heavy neutrinos to be hierarchical.

In next section we give the recent neutrino oscillation parameters. In section III we give a brief introduction to Type I seesaw and give the expressions for CP asymmetry for the different neutrino mass hierarchies and discuss each case explicitly in flavoured leptogenesis scenario. Finally in section IV we conclude with summary and discussion.
\section{Neutrino oscillation parameters}
The best fit and $1\sigma$ values of the mass squared differences and the mixing angles given by the recent neutrino oscillation data \cite{GonzalezGarcia:2012sz} are:
\bea
(\Delta m_{21}^2)_{bf}= 7.50\times 10^{-5} eV^2,\,\,\, 
7.31 \times 10^{-5}\leq \Delta m_{21}^2 \leq 7.68 \times 10^{-5} eV^2 \\
(\Delta m_{31}^2)_{bf}= 2.473\times 10^{-3} eV^2 \text{(NH)},\,\,
2.40 \times 10^{-3}\leq \Delta m_{31}^2 \leq 2.54 \times 10^{-3} eV^2\\
(\Delta m_{23}^2)_{bf}= 2.427\times 10^{-3} eV^2 \text{(IH)},\,\,\,
2.39\times 10^{-3}\leq \Delta m_{23}^2 \leq 2.49\times 10^{-3} eV^2\\
(\sin^2{\theta}_{12})_{bf}=0.302,\,\,\, 0.290\leq\sin^2{\theta}_{12} \leq 0.315 \\
(\sin^2{\theta}_{23})_{bf}=0.413,\,\, 0.388\leq\sin^2{\theta}_{23} \leq 0.450\\
(\sin^2{\theta}_{13})_{bf}=0.0227,\,\,0.020\leq\sin^2{\theta}_{13}\leq 0.025.
\eea
The above data gives three possible hierarchies of neutrino mass spectrum:\\
\bea \nonumber
 Normal\,\,Hierarchy \,\,:\,\, m_1 \ll m_2 <m_3. \\\nonumber
 Inverted\,\, Hierarchy\,\, :\,\,m_3 \ll m_1< m_2. \\ \nonumber
 Quasi-degenerate \,\,:\,\, m_1\simeq m_2\simeq m_3\simeq m.
\eea
 The $\Delta m_{21}^2$ denotes the solar mass squared difference $\Delta m_{\bigodot}^2$ while $\Delta m_{31}^2$ and $\Delta m_{23}^2$  denote the atmospheric mass squared differences $\Delta m_{atm}^2$ for normal and inverted hierarchy respectively. The leptonic mixing matrix is generally parametrised as:
\be
 U_{PMNS}=
 \left(\begin{array}{ccc}
 c_{13}c_{12}  &  c_{13}s_{12}  &   s_{13}\,e^{-i\delta}\\
 -c_{23}s_{12}-c_{12}s_{13}s_{23}\,e^{i\delta}   & c_{12}c_{23}-s_{12}s_{13}s_{23}\,e^{i\delta}  &   c_{13}s_{23} \\
 s_{12}s_{23}-c_{12}s_{13}c_{23}\,e^{i\delta}  &  -c_{12}s_{23}-c_{23}s_{13}s_{12}\,e^{i\delta}  &  c_{13}c_{23}
 \end{array}\right)\,diag\left(1,e^{i\alpha_1},e^{i\alpha_2}\right),
 \ee
where $s_{ij}=\sin\theta_{ij}$ and $c_{ij}=\cos\theta_{ij}$ respectively. Here $\delta$ is the Dirac CP phase and the Majorana phases are given by $\alpha_1$ and $\alpha_2$. Although the present oscillation data gives  very precise values of the mass squared differences and mixing angles but there are still no bounds on the Dirac and Majorana phases from the experiments. Moreover we do not have information about the exact value of the absolute neutrino mass and also the correct neutrino mass pattern.
\section{Type I seesaw Model and leptogenesis}
In the Type I seesaw addition of three RH neutrinos to the Standard Model(SM), which are singlet under the group $SU(2)\times U(1)$, gives the $3\times3$ light neutrino mass matrix to be
\be
m_{\nu}= m_{D}^T M_R^{-1}m_{D},
\ee 
where $m_{\nu}$ is the light neutrino mass matrix, $M_R$ is the right handed(RH) neutrino mass matrix and $m_D$ is the Dirac neutrino mass matrix. 
In the basis where the charged lepton Yukawa coupling are diagonal, the relevant terms in the lagrangian that gives the above seesaw formula after spontaneous symmetry breaking is :
\be 
{\cal L}_m = -\,\frac{1}{2}\,m_L{\nu}^T_L C^{\dag}{\nu}_L\, - \, m_D \nu_R \nu_L \,-\,\frac{1}{2}\,M_R{\nu}^T_R C^{\dag}{\nu}_R\  + h.c.
\ee
The first term in the above equation goes to zero as it is not invariant under $SU(2)\times U(1)$. Therefore, we have
\be
{\cal L}_m = - N_{L}^T C^{\dag}{\cal M} N_L + h.c,
\ee
where \be N_L = \left(\begin{array}{c} \nu_L \\ \nu_R^C \end{array}\right)\,\, {\text and}\,\,{\cal M} = \left(\begin{array}{cc}
  0 & m_D^T \\
 M_D & M_R \end{array}\right).
\ee 
For $ M_R \gg m_D$ the above mass matrix can be block diagonalised and we get the effective light neutrino mass matrix to be 
\be
m_{\nu}\simeq m_{D}^T M_R^{-1}m_{D} = v^2 Y_{\nu}^T M_R^{-1}Y_{\nu}, 
\ee
where v is the vev and $Y_{\nu}$ is the neutrino Yukawa coupling matrix. Here we consider the RH Majorana mass matrices to be real and diagonal such that $M_R = M_R^{dia}=Dia\left(M_1,M_2,M_3\right)$. Under such an assumption the low energy phases  in $m_{\nu}$ can appear only in the Yukawa couplings. The PMNS matrix diagonalising $m_{\nu}$ is given by
\be
U^Tm_{\nu}U=dia\left(m_1,m_2,m_3\right)=m^{dia}.
\ee
Using the Casas Ibarra parametrisation \cite{Casas:2001sr} the  $Y_{\nu}$'s 
can be written as
\be  \nonumber
Y_{\nu}= \frac{1}{v}\sqrt{M_R^{dia}}R\sqrt{m^{dia}} U^{\dag}.
\ee
For a particular lepton flavour $l$ where $l=e,\,\mu,\,\tau$, 
\be 
Y_{i\,l} =  \frac{1}{v}\sqrt{M_i}R_{ik}\sqrt{m_k} U_{l\, k}^{*}.
\ee 
The matrix $R$ is a orthogonal matrix which is in general complex but in our 
case it is a real matrix as  CP is an exact symmetry of the RH sector. 
The self energy and vertex corrections of the decays : $N_1\rightarrow l+ \phi$ and  $N_1\rightarrow {\bar l} + \phi^{\dag}$ gives a CP asymmetry because of the difference in the decay rates of the two modes. Taking flavour effects \cite{Abada:2006ea} into account the lepton asymmetry for each flavour $l$ is given by 
\be
\epsilon_{l}= -\frac{3M_1}{16 \pi v^2}\,\frac{Im \left(\sum_{\alpha\beta}m_\al^{1/2}m_{\beta}^{3/2}\,U^*_{l\al}U_{l\beta}R_{1\al}R_{1\beta}\right)}{\sum_{\al} m_{\al} |R_{1\al}|^2}
\label{cpflavour}
\ee 
and the total CP asymmetry $\epsilon_1 =\sum_{l}\epsilon_{l}$. Since $U$ is a unitary matrix and R is real here, we get the total asymmetry $\epsilon_1$ to be zero when summed over all the lepton flavours. We should note here that for the CP asymmetry in each flavour to be non-zero it also requires $R$ to be non-diagonal. It shows that single flavour approximation models with exact CP symmetry in the RH neutrino sector gives vanishing lepton asymmetry. But we have already mentioned that single flavour approximation is true only for temperature greater than $10^{12}\, GeV$.  
If we go to temperatures  $T\leq 10^{12}\,GeV$ the tau lepton Yukawa interactions  come into equilibrium and if we go below $10^{9}\,GeV$, the ${\mu}$ Yukawa couplings also come into equilibrium. It is in this temperature region  $ 10^{9}\leq T \leq 10^{12}\,GeV$ where the tau leptons becomes distinguishable from the $e$ and $\mu$. Then the total CP asymmetry, which is the sum of CP asymmetry due to $\tau$ and $e\,+\,\mu$ is non zero. We prefer to work in the range $ 10^{9}\leq T \leq 10^{12}\,GeV$ in our analysis, where only $\tau$ leptons are in equilibrium. 
The baryon asymmetry also depends on the wash out parameter, which for each flavour $l=e,\mu,\tau$ is given by
\be
{\tilde m}_l\equiv \frac{Y_{l1}^*\,Y_{l1}\,v^2}{M_1}=\sum_{i}R_{1j}^2m_{j}^2U_{lj}^*U_{lj}, \,\,\,{\text j=1,2,3}
\label{wash-out}
\ee
and 
\be
m_*\equiv 8\pi\frac{v^2}{M_1^2}H|_{T=M_1}\simeq 1.1\times10^{-3} eV,
\label{mstar}
\ee
where the ${\tilde m}_l$ and $m_*$ are related to the decay rate of the RH neutrino $N_1$ and the expansion rate of the universe respectively. Therefore the  efficiency factor is
\be
\eta({\tilde m}_l) \simeq \left(\left(\frac{\tilde m_l}{8.25\times 10^{-3} eV} \right)^{-1}\,+\, \left(\frac{0.2\times10^{-3}}{\tilde m_l}\right)^{-1.16}\right)^{-1}.
\label{efficiency}
\ee
The final baryon asymmetry which is the sum of the asymmetries in each flavour, can be obtained after solving the Boltzmann equations taking flavour effects into account \cite{Abada:2006ea}: 
\be
Y_{B}\simeq -\frac{12}{37\,g^*}\left(\epsilon_{2}\eta\left(\frac{417}{589}{\tilde m_2}\right)\,+\,\epsilon_{\tau}\eta\left(\frac{390}{589}{\tilde m_{\tau}}\right)\right).
\label{finalYB}
\ee
As the lepton asymmetries in $e$ and $\mu$ are indistinguishable in this range, we can combine the two lepton asymmetries and the wash out parameters such that 
\be
\epsilon_{2}=\epsilon_{e}\,+\,\epsilon_{\mu}=-\epsilon_{\tau},\,\,\,\,{\tilde m_2}={\tilde m_e}\,+\,{\tilde m_{\mu}}.
\label{mteqm2}
\ee
Moreover, the above equation shows that it is sufficient to calculate the CP asymmetry of the $\tau$ in this particular limit. Throughout the analysis we have also neglected the scatterings by heavier RH neutrinos $N_{2,3}$.  
\subsection{Leptogenesis in $\nu$ mass models with Normal hierarchy ($m_1\ll m_2\ll m_3$):}
In this section we consider leptogenesis in light neutrino mass models with Normal hierarchy mass pattern such that $m_2\simeq \sqrt{\Delta m_{\bigodot}^2}$ and 
$m_3\simeq  \sqrt{\Delta m_{atm}^2}$. As $m_1 \ll m_{2,3}$ the CP asymmetry term in eqn.(\ref{cpflavour}) gives
\bea \nonumber
\epsilon_{l}&=&-\frac{3M_1}{16 \pi v^2}\,\left( \frac{m_2^{1/2}m_{3}^{3/2}R_{12}R_{13}\,Im \left( U^*_{l2}U_{l3}\right)}{ m_{2}R_{12}^2\,+\,m_3R_{13}^3} \,+\,
 \frac{m_3^{1/2}m_{2}^{3/2}R_{12}R_{13}\,Im \left( U^*_{l2}U_{l3}\right)}{ m_{2}R_{12}^2\,+\,m_3R_{13}^3} \right)\\ \nonumber
&=& -\frac{3M_1}{16 \pi v^2}\,\frac{(\Delta m_{\bigodot}^2\Delta m_{atm}^2)^{1/4}\,(1-\rho)\,R_{12}R_{13}}{\rho R_{12}^2 \,+\, R_{13}^2}\,Im \left( U^*_{l2}U_{l3}\right), \,\,{\text where}\, \rho = \sqrt{\frac{\Delta m_{\bigodot}^2}{\Delta m_{atm}^2} }, \\
&=& -\frac{3M_1}{16 \pi v^2}\,\frac{\sqrt{\Delta m_{atm}^2}\sqrt{\rho}\left( 1-\rho \right)R_{12} R_{13}}{\rho R_{12}^2 \,+\, R_{13}^2}\,Im \left( U^*_{l2}U_{l3}\right). 
\eea

Therefore the CP asymmetry $\epsilon_{\tau}$ is
\bea \nonumber
\epsilon_{\tau}= -\frac{3M_1}{16 \pi v^2}\,\frac{\sqrt{\Delta m_{atm}^2}\sqrt{\rho}\left( 1-\rho \right) R_{12} R_{13}}{\rho R_{12}^2 \,+\, R_{13}^2}\,\times 
c_{13} c_{23} 
 \left\{c_{12} s_{23} \sin\left(\frac{\alpha_1 - \alpha_2}{2}\right)\right. \\
+ \left. 
   c_{23} s_{12} s_{13} \sin\left(\frac{\alpha_1 - \alpha_2}{2}\,+\, \delta \right) \right\}.
\label{epsilon-nh}
\eea
Further we get
\bea \nonumber 
{\tilde m}_2= \sqrt{\Delta m_{atm}^2}\left\{ R_{13}^2 \left(s_{13}^2 + c_{13}^2 s_{23}^2\right)+\, R_{12}^2\rho\left[c_{12}^2 c_{23}^2 +s_{12}^2 c_{13}^2+s_{12}^2 s_{13}^2 s_{23}^2 - \frac{1}{2} \sin 2\theta_{12} \sin 2\theta_{23} s_{13} \cos\delta\right]\right.\\ \nonumber
+ \left. \, R_{12} R_{13}\sqrt{\rho} \left[c_{12} c_{13}\sin2\theta_{23}\cos\left(\frac{\alpha_1 - \alpha_2}{2}\right)\,+\, s_{12} \sin 2 {\theta}_{13}\, c_{23}^2 \cos \left(\frac{\alpha_1 - \alpha_2}{2}+\delta \right)
\right]  
 \right\}
\label{m2nh}
\eea
and
\bea  
{\tilde m}_{\tau}= \sqrt{\Delta m_{atm}^2}
\left\{
R_{13}^2c_{13}^2c_{23}^2 + R_{12}^2\rho\left[c_{23}^2s_{12}^2s_{13}^2+ c_{12}^2s_{23}^2 + \frac{1}{2} sin2\theta_{12}sin2\theta_{23}s_{13}cos\delta \right]\right.\\
\left. -R_{12}R_{13}\sqrt{\rho}\left[ c_{12}c_{13}sin2\theta_{23}\cos\left(\frac{\alpha_1 - \alpha_2}{2}\right)+c_{23}^2s_{12}sin2\theta_{13} \cos \left(\frac{\alpha_1 - \alpha_2}{2}+\delta \right)\right]
\right\}.
\label{mtnh}
\eea
Using eqn.(\ref{mteqm2}) in eqn.(\ref{finalYB}) we get
\be
Y_B\simeq -\frac{12}{37}\frac{\epsilon_{\tau}}{g^*}\,\left( \eta\left(\frac{390}{589}{\tilde m_{\tau}}\right)\,+\,\eta\left(\frac{417}{589}{\tilde m_2}\right)
\right).
\label{YBnh}
\ee
\subsubsection{Values of $R_{12}$ and $R_{13}$ for strong wash-out }
The condition for strong wash-out is $\tilde m_{\tau}\gg m^*$. We use eqn.(\ref{mstar}) and get the bound on the value of $R_{12}$ and $R_{13}$ to have a strong wash out
\be \nonumber
R_{12}^2 \gg \frac{1.1\times 10^{-3} eV}
{\sqrt{\Delta m_{\bigodot}^2}\left[ c_{23}^2s_{12}^2s_{13}^2\,+\,c_{12}^2s_{23}^2\,+\, \frac{1}{2}sin 2 \theta_{12} sin 2 \theta_{23}s_{13}cos\delta\right]}. 
\ee
The dominating term in the denominator of the above equation is $c_{12}^2s_{23}^2$. For the central value of the oscillation parameters, the term $c_{23}^2s_{12}^2s_{13}^2$ is of the order of ${\cal O}(100)$ smaller then the dominating term due to the presence of $s_{13}^2$. 
The term containing $cos\delta$ can be either positive or negative depending upon whether we are taking $\delta$ to be maximum(zero) or minimum($\pi$). Thus we find the approximate lower bound for $R_{12}$ to be 
\be
R_{12}\gg 0.75\,\,\,\,\, {\text for}\,\,\, \delta= 0 \,\,\, 
{\text and}\,\,\,R_{12} \gg 0.59\,\,\,\, {\text for}\, \delta= \pi 
\ee
for the strong wash-out regime and similarly for $R_{13}$ we get 
\be
R_{13}^2 \gg \frac{1.1\times 10^{-3} eV}{\sqrt{\Delta m_{atm}^2}\,c_{12}^2c_{23}^2}\,\,\,\,\Rightarrow R_{13}\gg 0.2.
\ee 
For the calculations in the strong wash-out regime we use $R_{12}=0.81$, $R_{13}=0.5$ and $M_1=5.0\times10^{11}\, GeV$. Now using eqn.(\ref{epsilon-nh}), eqn.(\ref{m2nh}), eqn.(\ref{mtnh}) and the $1\sigma$ values of the neutrino oscillation parameters and putting these into the eqn.(\ref{YBnh}) we try to fix the allowed low energy phases for the observed $1\sigma$ range of $Y_B$$(8.294\times10^{-11} - 8.508\times10^{-11})$. 
\begin{figure}
\begin{center}
\subfigure[]{
\includegraphics[width=7.5cm,height=7.0cm]{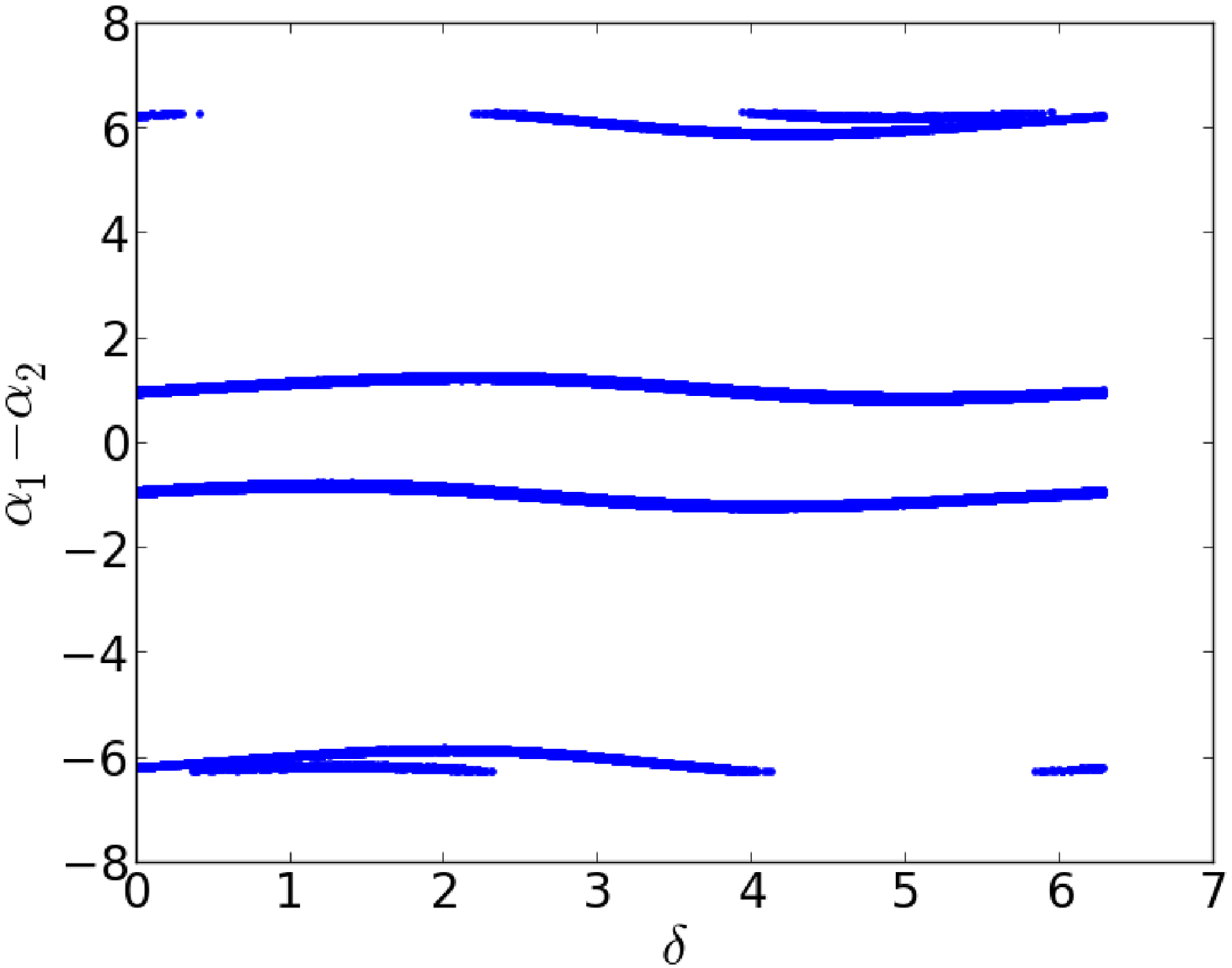}
}
\subfigure[]{
\includegraphics[width=7.5cm,height=7.0cm]{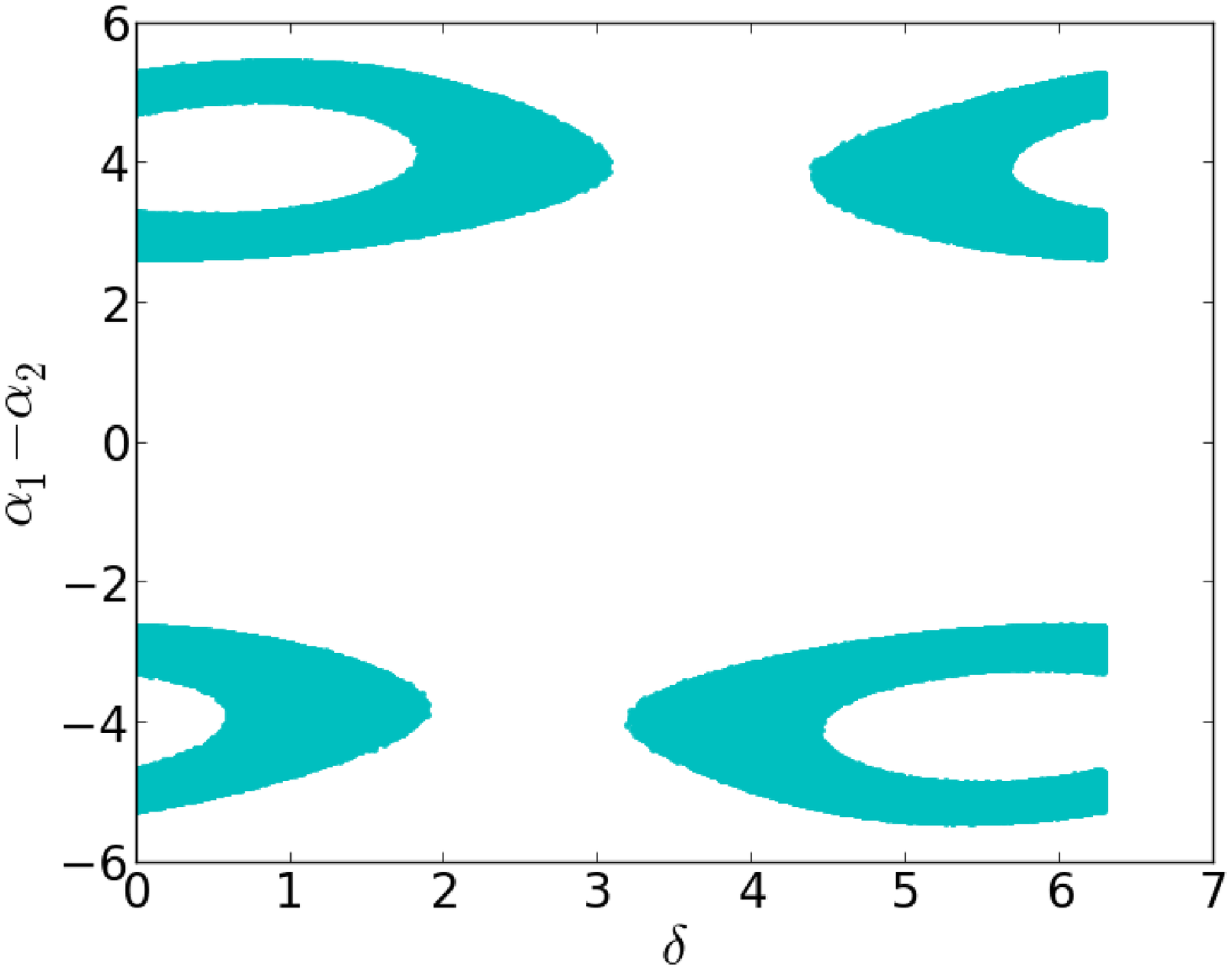}
}
\caption{The left panel and the right panel show the allowed low energy Dirac and Majorana phases for the observed $1\sigma$ range of BAU in normal hierarchical neutrino mass model in strong wash-out and weak wash-out regime respectively. The figures are for $1\sigma$ value of the oscillation parameters and $M_1=5.0\times10^{11}\, GeV$.}
\label{fig:strong_nh}
\end{center}
\end{figure}
The left panel of fig.\ref{fig:strong_nh} shows the correlation between the effective Majorana phases $(\alpha_1-\alpha_2)$ and Dirac CP phase $\delta$ which appears in the expression for CP asymmetry in the strong wash out regime in normal hierarchy. We can see from the fig.\ref{fig:strong_nh} that for $\delta$ varying from 0 to $2\pi$ only certain regions of $(\alpha_1-\alpha_2)$ are  allowed for the observed range $1\sigma$ of baryon asymmetry. This is different from the earlier analysis \cite{Branco:2006ce} where the authors did the analysis for the best-fit value of the neutrino oscillation parameters. In the present work we scan over the $1\sigma$ range of the recent neutrino oscillation data and try to find the allowed values of Dirac and Majorana phase which can generate the required baryon asymmetry. The study becomes important with the recent measurement of the neutrino reactor angle $\theta_{13}$ and also with the updated measurement of the BAU.    

\subsubsection{Values of $R_{12}$ and $R_{13}$ for weak wash-out }
The condition for weak wash-out $(\tilde m_{\tau} \ll m^*)$ requires the value of $R_{12}\ll 0.75$ and $R_{13}\ll 0.2$. So we can safely consider $R_{12}=0.45$ and $R_{13}=0.01$ for our analysis in the weak-wash out regime. The right panel of fig.\ref{fig:strong_nh} shows the correlation of the phases $(\alpha_1-\alpha_2)$ and $\delta$ in the weak wash out regime for normal hierarchy mass model for the above values of $R_{12}$ and $R_{13}$. Here too we see that for all the values of $\delta$ i.e; from zero to $2\pi$, only certain values of $(\alpha_1-\alpha_2)$ are allowed and most of the region are excluded.
\begin{figure}[ht]
\begin{center}
\includegraphics[width=8.5cm,height=6cm]{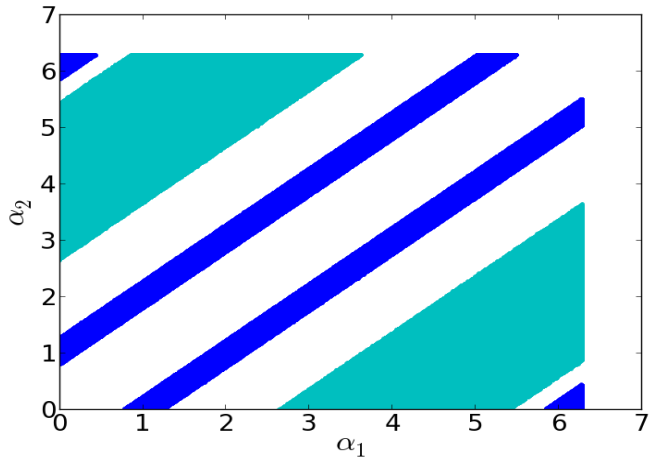}
\caption{The figure shows the allowed Majorana phases in strong(blue) and weak(cyan) wash out regime for neutrino mass model with normal hierarchy. The plots are for $1\sigma$ value of the $\nu$ oscillation parameters and $M_1=5.0\times10^{11}\, GeV$.}
\label{fig:strong_weak}
\end{center}
\end{figure}
Fig.\ref{fig:strong_weak} depicts the relation between Majorana phases in strong and weak wash out regime denoted by blue and cyan colours respectively. It is clear from the figure that some of the regions are totally excluded both in weak as well in strong wash out. 

\begin{figure}[ht]
\begin{center}
\subfigure[]{
\includegraphics[width=7.5cm,height=6.2cm]{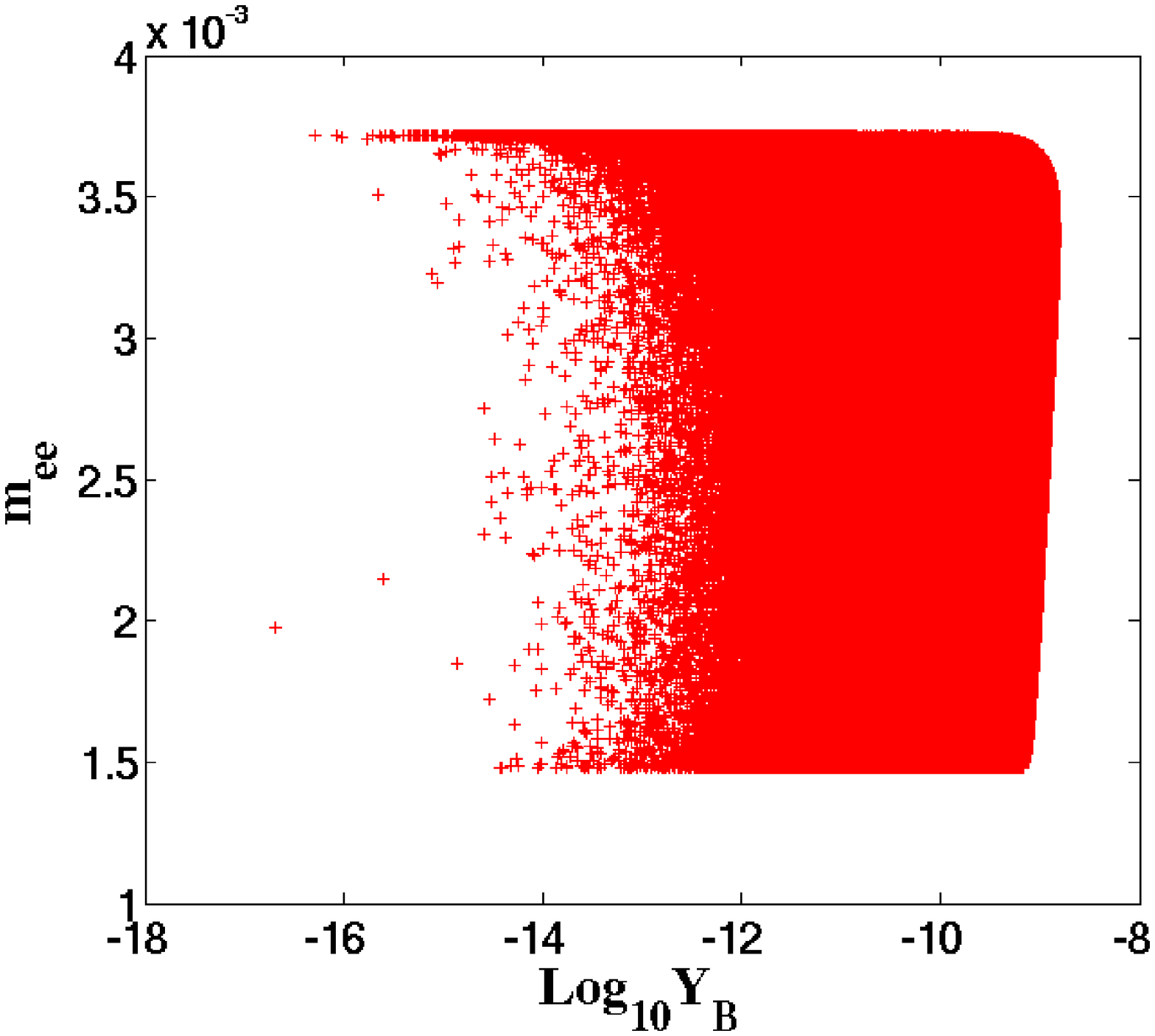}
}
\subfigure[]{
\includegraphics[width=7.5cm,height=6.0cm]{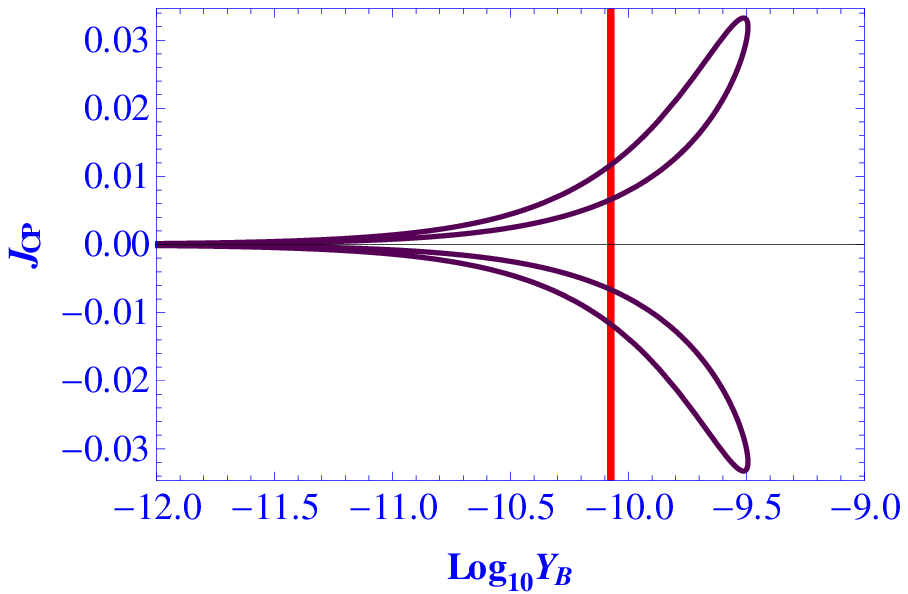}
}
\caption{ The left panel shows $Y_B$ vs. $|m_{ee}|$ for $\delta=0$ and $\alpha_1-\alpha_2$ varying from 0 to $2\pi$. We take $R_{12}=0.81$ and $R_{13}=0.5$. The right panel shows $Y_B$ vs. $J_{CP}$ for $\delta$ varying from 0 to $2\pi$, $\alpha_1-\alpha_2=0$ and best-fit values of the neutrino oscillation parameters. The red line shows the recent $1\sigma$ observed value of $Y_B$.}
\label{fig:yB}
\end{center}
\end{figure}
The left panel of fig.\ref{fig:yB} shows the relation between $Y_B$ and the effective Majorana mass when the CP violating Dirac phase $\delta$ is set to zero. The range of the effective Majorana mass $|m_{ee}|$ in the allowed range of $Y_B$ is far below the experimental reach of GERDA \cite{Schonert:2005zn} which is 10 meV. But we can relate the Jarkslog Invariant $J_{CP}$ to $Y_B$ as shown  in the right panel of fig.\ref{fig:yB} setting the Majorana phases to zero. 
 Although this was studied by Pascoli et al.\cite{Pascoli:2006ci}, we find a significant change in the allowed values of $J_{CP}$ with the present values of $Y_B$ and recent oscillation parameters. These allowed  values now lie between 
-0.01 to -0.006 and 0.006 to 0.01 for the best-fit values of the neutrino oscillation parameters, whereas the earlier allowed values of $J_{CP}$ were lying between 0.02 to 0.03 and -0.03 to -0.02.   
\subsection{Leptogenesis in $\nu$ mass models with Inverted hierarchy ($m_2>m_1\gg m_3$):}
In the inverted hierarchy model we consider the case where the contributions from the terms containing $m_3$ are negligible  as compared to $m_1$ and $m_2$. Here, the lepton flavour CP asymmetry expression under this approximation looks like
\bea \nonumber
\epsilon_{l}&=&-\frac{3M_1}{16 \pi v^2}\,\left( \frac{m_1^{1/2}m_{2}^{3/2}R_{11}R_{12}\,Im \left( U^*_{l1}U_{l2}\right)}{ m_{1}R_{11}^2\,+\,m_2R_{12}^3} \,+\,
 \frac{m_2^{1/2}m_{1}^{3/2}R_{12}R_{11}\,Im \left( U^*_{l2}U_{l1}\right)}{ m_{1}R_{11}^2\,+\,m_2R_{12}^3} \right)\\ \nonumber
&=& -\frac{3M_1}{16 \pi v^2}\,\frac{m_1^{1/2}m_2^{1/2}}{ m_{1}R_{11}^2\,+\,m_2R_{12}^3}\left[m_2Im\left(U^*_{l1}U_{l2}\right) \,+\, m_2Im\left(U_{l2}^*U_{l1}\right)\right]
\eea
We have $m_2\simeq m_1\simeq \sqrt{\Delta m_{atm}^2}$ and from the relations $m_2^2-m_1^2=\Delta m_{\bigodot}^2$ and $m_2^2-m_3^2= \Delta m_{atm}^2$ we get  $m_2-m_1 \simeq \frac{\Delta m_{\bigodot}^2}{2\,\Delta m_{atm}^2}$. Substituting all these in the above expression we get
\be \nonumber
\epsilon_{l}=-\frac{3M_1}{32 \pi v^2}\,\frac{R_{11}R_{12}}{R_{11}^2 + R_{12}^2}\,\sqrt{\Delta m_{\bigodot}^2}\,\rho\,Im\left(U_{l1}^*U_{l2}\right),
\ee
\bea \nonumber
\simeq\,-\frac{3M_1}{32 \pi v^2}\,\frac{R_{11}R_{12}}{R_{11}^2 + R_{12}^2}\,\sqrt{\Delta m_{\bigodot}^2}\,\rho\,\times\,\frac{1}{2}c_{12}s_{12}\left(c_{23}^2s_{13}^2-s_{23}^2\right)\sin\frac{\alpha_1}{2}\\  \nonumber
 +\,\frac{1}{2}\sin 2\theta_{23}s_{13}\left[ c_{12}^2\sin(\frac{\alpha_1}{2}-\delta)\,-\, s_{12}^2\sin(\frac{\alpha_1}{2}+\delta) \right],\\ 
\simeq\,-\frac{3M_1}{32 \pi v^2}\,\frac{R_{11}R_{12}}{R_{11}^2 + R_{12}^2}\,\sqrt{\Delta m_{\bigodot}^2}\,\rho\,\times\,\frac{1}{2}c_{12}s_{12}\left(c_{23}^2s_{13}^2-s_{23}^2\right)\sin\frac{\alpha_1}{2}\\ \nonumber 
+\, \frac{1}{2}\sin 2\theta_{23}s_{13}\left(\cos\delta\sin\frac{\alpha_1}{2}\cos 2\theta_{12}\,-\,\cos\frac{\alpha_1}{2}\sin \delta \right).
\eea
The wash-out factors ${\tilde m}_{2\tau}$ are
\bea \nonumber
{\tilde m}_2 &\simeq&\sqrt{\Delta m_{atm}^2}\,
\left\{  
R_{12}^2c_{12}^2c_{23}^2 \,+\, R_{12}^2s_{12}^2c_{13}^2\,+\, R_{12}^2s_{12}^2s_{13}^2s_{23}^2 
\,-\,\frac{1}{2}R_{12}^2\sin 2{\theta}_{12}\,\sin{2\theta}_{23}s_{13}\cos\delta
\,+ \, R_{11}^2c_{23}^2s_{12}^2 \right.\\ \nonumber
&&\left.\,+\, R_{11}^2c_{12}^2c_{13}^2 \,+\,R_{11}^2c_{12}^2s_{13}^2s_{23}^2 \,-\, \frac{1}{2}R_{11}^2\sin 2\theta_{12}
\sin 2\theta_{23}s_{13}\cos\delta\,+\, R_{11}R_{12}sin 2\theta_{12}c_{12}^2
s_{23}^2\cos\frac{\alpha_1}{2} \right.\\  
&&\left.\,-\, R_{11}R_{12}\sin 2\theta_{23}c_{12}^2s_{13}
cos(\frac{\alpha_1}{2} -\delta)\,+\, R_{11}R_{12} \sin 2\theta_{23}s_{12}^2s_{12}^2
s_{13}\cos(\frac{\alpha_1}{2}-\delta)\right\}.
\eea
\begin{figure}
\subfigure[]{
\includegraphics[width=7.4cm,height=6.0cm,trim = 0 0 0 2]{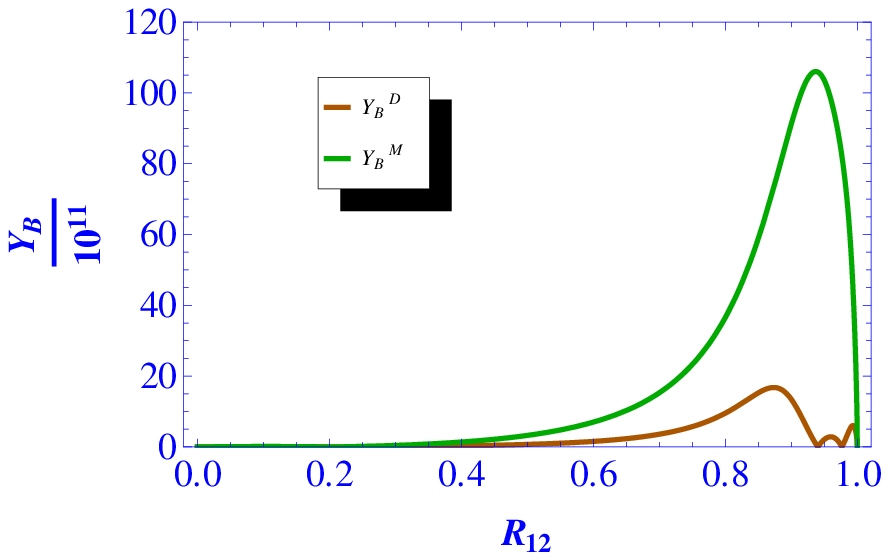}
}
\subfigure[]{
\includegraphics[width=7.6cm,height=6.3cm,trim = 0 0 0 2]{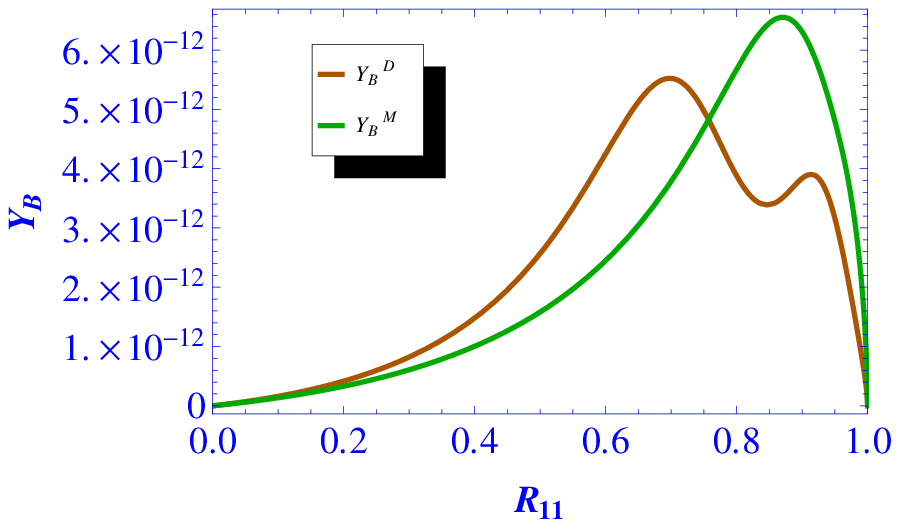}
}
\caption{The left panel and the right shows the BAU generated for normal hierarchy and inverted hierarchy model respectively for best fit values of the oscillation parameters and for $R_{12}$ and $R_{11}$ ranging from 0 to 1. The thick green curve is for only Majorana phases contribution and the dashed red curve denotes only Dirac phase contribution to $Y_B$.}
\label{fig:lepto-nh}
\end{figure}
\bea \nonumber
{\tilde m}_{\tau}&\simeq&\sqrt{\Delta m_{atm}^2}\,
\left\{  c_{23}^2s_{12}^2s_{13}^2R_{12}^2\,+\, c_{12}^2S_{23}^2R_{12}^2\,+\,R_{12}^2\sin 2\theta_{12}\sin 2\theta_{23}s_{13}cos\delta \right.\\ \nonumber
&&\left.+\, R_{11}^2 \left( c_{12}^2s_{13}^2c_{23}^2\,+\, s_{12}^2s_{23}^2\,-\, \frac{1}{2}\sin 2theta_{12}\sin 2 \theta_{23}s_{13}cos\delta \right)\right.\\ \nonumber
&& \left.+\, R_{11}R_{12}\sin 2\theta_{12}\left(c_{23}^2s_{13}^2-s_{23}^2\right)\cos\frac{\alpha_1}{2}
\,+\,R_{11}R_{12}\sin2\theta_{23}c_{12}^2s_{13}cos\left(\frac{\alpha_1}{2}-\delta \right) \right.\\ 
&&\left.-\, R_{11}R_{12}\sin 2\theta_{23}s_{12}^2s_{13}\cos\left(\frac{\alpha_1}{2}+\delta \right)\right\}.
\eea
We have ${\tilde m}_2 \,+\,{\tilde m}_{\tau}=\sqrt{\Delta m_{atm}^2}$.  In this  particular case of the heaviest RH neutrino $N_3$ decoupling, we have $R_{11}^2 +R_{12}^2=1$ \cite{Petcov:2005jh}. Therefore we vary the values of $R_{11}^2$ and $R_{12}^2$ between 0 and 1. The left and the right panel of fig.\ref{fig:lepto-nh} shows the baryon asymmetry that can be generated for normal and inverted hierarchy respectively. The thick green colour curves are for $Y_B$ generated with only Majorana phases and the dashed red colour curves denote $Y_B$ only via Dirac CP phase. In the fig.\ref{fig:lepto-nh} we have put the figure for normal hierarchy for the sake of comparison with the inverted hierarchy. As we can see that the normal hierarchy model can generate the required  $Y_B$ for the Majorana phase contribution but the inverted hierarchy model can generate a maximum  $Y_B$$\sim6.0\times10^{-12}$. This is far below the observed value and therefore it is not possible to generate the observed baryon asymmetry in the inverted hierarchy neutrino mass models with CP symmetric RH sector. This is because in this case the CP symmetry term is suppressed by the term $\sqrt{\Delta m_{\bigodot}^2}\times \rho$. Our result for the inverted hierarchy agrees with the earlier works \cite{Branco:2006ce,Pascoli:2006ci}. 
\subsection{Leptogenesis in Quasi-degenerate models  ($m=m_1\simeq m_2\simeq m_3\gg \sqrt{\Delta m_{atm}^2}$):}
For the quasi-degenerate models the CP asymmetry term is given by:
\bea \nonumber
\epsilon_{\tau}&=&\frac{3M_1}{16 \pi v^2}\frac{1}{\sum_im_iR_{1i}^2}\,Im \left\{R_{11}^2m_1|U_{\tau 1}|^2 \,+\, R_{11}R_{12}m_1^{1/2}m_2^{3/2}U_{\tau 1}^*U_{\tau 2}\,+\, R_{11}R_{13}m_1^{1/2}m_3^{3/2}U_{\tau 1}^*U_{\tau 3}\, \right.\\ \nonumber
&&\left.+\,R_{11}R_{12}m_2^{1/2}m_1^{3/2}U_{\tau 2}^*U_{\tau 1}\,+\,R_{22}^2m_1|U_{\tau 2}|^2 \,+\, R_{12}R_{13}m_2^{1/2}m_3^{3/2}U_{\tau 2}^*U_{\tau 3} 
\right.\\ \nonumber
&&\left. R_{11}R_{13}m_3^{1/2}m_1^{3/2}U_{\tau 3}^*U_{\tau 1}\,+\,R_{13}R_{12}m_3^{1/2}m_2^{3/2}U_{\tau 3}^*U_{\tau 2}\,+\, ,R_{33}^2m_3|U_{\tau 3}|^2 
 \right\} \\ \nonumber
&=&\frac{3M_1}{16 \pi v^2}\frac{1}{\sum_im_iR_{1i}^2}\,\left\{R_{11}m_1|U_{\tau 1}|^2 \,+\, R_{22}m_2|U_{\tau 2}|^2 +R_{33}m_3|U_{\tau 3}|^2  \right.\\ \nonumber
&&\left. +\, R_{11}R_{12}m_1^{1/2}m_2^{1/2}(m_2-m_1)Im\left(U_{\tau 1}^*U_{\tau 2}\right)  +\, R_{11}R_{13}m_1^{1/2}m_3^{1/2}(m_3-m_1)Im\left(U_{\tau 1}^*U_{\tau 3}\right) \right.\\ \nonumber 
&&\left. +\, R_{12}R_{13}m_2^{1/2}m_3^{1/2}(m_3-m_2)Im\left(U_{\tau 2}^*U_{\tau 3}\right) 
\right\}  \\ \nonumber
&=& \frac{3M_1}{16 \pi v^2}\frac{1}{\sum_{i}R_{1i}^2}\left\{  R_{11}R_{12}(m_2-m_1)Im\left(U_{\tau 1}^*U_{\tau 2}\right)\,+\, R_{11}R_{13}(m_3-m_1)Im\left(U_{\tau 1}^*U_{\tau 3}\right)\right. \\ \nonumber
&&\left.\,+\,R_{12}R_{13}(m_3-m_2)Im\left(U_{\tau 2}^*U_{\tau 3}\right) \right\}.
\eea
Ignoring the terms $m_2-m_1\sim  \frac{\sqrt{\Delta m_{\bigodot}^2}}{2 m} $, $m_3-m_1\simeq\frac{\Delta m_{atm}^2}{2 m}$ and $m_3-m_2\simeq\frac{\Delta m_{atm}^2}{2 m}$ we get
\bea  \nonumber
\epsilon_{\tau}&\simeq&\frac{3M_1}{16 \pi v^2}\frac{\Delta m_{atm}^2}{2 m}\times \left\{  R_{11}R_{13}Im\left(U_{\tau 1}^*U_{\tau 3}\right) + R_{12}R_{13}Im\left(U_{\tau 2}^*U_{\tau 3}\right) \right\}\\ \nonumber
&=& \frac{3M_1}{32 \pi v^2}\frac{\Delta m_{atm}^2}{m}\times 
\left\{  \frac{1}{2}R_{11}R_{13}\left( c_{13}\sin 2\theta_{23}s_{12}\sin\frac{\alpha_2}{2} -\sin 2 \theta_{13}c_{23}^2c_{12}\sin(\frac{\beta}{2}-\delta)\right)\right.\\ \nonumber
&&\left. \frac{1}{2}R_{12}R_{13}\left( c_{13}\sin 2\theta_{23}c_{12}\sin
\frac{\alpha_1-\alpha_2}{2}+ c_{23}^2s_{12}\sin 2 \theta_{12}
\sin\left(\frac{\alpha_1-\alpha_2}{2}+\delta \right)\right)\right\}\\ \nonumber
&=& \frac{3M_1}{64 \pi v^2}\frac{\Delta m_{atm}^2}{m}\times 
\left\{  R_{11}R_{13}\left[ c_{13}\sin 2\theta_{23}s_{12}\sin\frac{\alpha_2}{2} -\sin 2 \theta_{13}c_{23}^2c_{12}\sin\left(\frac{\beta}{2}-\delta\right)\right]\right.\\
&&\left. R_{12}R_{13}\left[ c_{13}\sin 2\theta_{23}c_{12}\sin\frac{\alpha_1-\alpha_2}{2}+ c_{23}^2s_{12}\sin 2 \theta_{12}\sin\left(\frac{\alpha_1-\alpha_2}{2}+\delta \right)\right]
\right\}.
\label{cptaudeg}
\eea
The wash out factor (see eqn.(\ref{wash-out})in the quasi-degenerate case becomes
\be
{\tilde m}_e \,+\,{\tilde m}_{\mu} \,+\,{\tilde m}_{\tau}= m\left(R_{11}^2 + R_{12}^2+R_{13}^2\right)
\ee
and
\be
{\tilde m}_l=m\left|R_{11}U_{l 1}^*+R_{12}U_{l 2}^*+R_{13}U_{l 3}^*\right|^2.
\label{mtaudeg}
\ee
In the degenerate case the wash out factor is proportional to the
 absolute neutrino mass $m$. So we can see from eqn.(\ref{efficiency}) that for smaller values of the neutrino absolute mass we can have large efficiency factors. Therefore for the strong wash out condition ${\tilde m}_{\tau}$ is much smaller than the absolute mass $m$ i.e, ${\tilde m}_{\tau}\ll m$. For the best-fit values of the neutrino oscillation parameters we get 
\be
{\tilde m}_{\tau}= m\,R_{13}^2\,\times 0.1889 
\ee
 for $\alpha_1=\alpha_2=\pi$ and $\delta=\pi/2$ and also taking $R_{11}=R_{12}=R_{13}$. Therefore we have
\be
R_{13}^2\gg \frac{1.1\times10^{-3}}{m\times 0.1889}
\ee
for strong wash-out.
The {\it Planck} measurements \cite{Ade:2013zuv} gives sum of the neutrino mass bound to be $\sum m_{\nu}< 0.23 \, eV$. For the smallest allowed value of absolute mass $m = 0.07\,eV$, we get $R_{13}^2\gg 0.13$. In the strong wash out regime considering $R_{13}^2=0.26$ we get the maximum efficiency to be
\bea \nonumber
\eta_{max}&\simeq &\left(\frac{0.2\times 10^{-3}}{\frac{390}{589}{\tilde m}_{\tau}}\right)^{1.16}\\
&=& 0.0595. 
\eea 
Using eqn.(\ref{cptaudeg}) one can obtain the CP asymmetry for the above values of  CP phases ($\alpha_1=\alpha_2=\pi$ and $\delta=\pi/2$) and taking $R_{11}=R_{12}=R_{13}$. 
Using eqn.(\ref{finalYB}) and $M_1= 5.0\times 10^{11}\,GeV$ we get the maximum value of the $Y_B$ to be less than $2.58\times 10^{-19}$ for lowest allowed value of the neutrino absolute mass $m$. This is much smaller than the observed value of $Y_B$.
 
Similarly for the weak wash-out condition we require ${\tilde m}_{\tau}\ll 0.13$. Considering $R_{13}^2 = 0.06$ the maximum efficiency is found to be $0.0638$ and $Y_B$ to be less than $2.76\times 10^{-19}$. Thus we see that inorder to achieve the range of observed BAU we need $M_1\gg 10^{12}\, GeV$ and a much lower value of the absolute neutrino mass. Therefore the Quasi-Degenerate neutrino mass model does not seem to be favourable choice if we have an exact CP symmetric and hierarchical heavy right handed neutrinos.
\section{Summary and Conclusion:} 
 Low energy CP violating phases responsible for low energy leptonic CP violation may or may not be responsible for CP violation in high energy scale. Therefore it is in general not possible to connect both the CP violations. But if we consider the right handed sector to be CP symmetric then  we have the advantage of generating a non-zero CP asymmetry from the low energy phases alone in flavoured leptogenesis scenario \cite{Branco:2006ce,Pascoli:2006ie}. Under this particular assumption only normal hierarchical light neutrino mass model is able to generate the required baryon asymmetry of the universe with the present values of the neutrino oscillation data. 

In this work we have analysed this particular scenario in details taking into account the recent neutrino oscillation data \cite{GonzalezGarcia:2012sz} and {\it Planck} results \cite{Ade:2013zuv}  and also considering the right handed heavy neutrinos to be hierarchical. We find that only certain combinations of effective Majorana and Dirac CP phases are allowed in normal hierarchy for the recent $1\sigma$ value of the baryon asymmetry $Y_B$. These combinations are different for the strong and weak wash-out regimes. Our analysis differs from the previous works, as we have varied the neutrino masses and mixing parameters  within $1\sigma$ range of recent oscillation data. We have also shown that in inverted hierarchy models the maximum BAU generated is about $6\times10^{-12}$ which is much less then the observed value of $8.40\times10^{-11}$ and is in agreement with the earlier works. Therefore within this particular scenario the inverted hierarchy neutrino mass model is not a favourable one. In the quasi-degenerate case also it is not possible to generate the observed $Y_B$ in the temperature region we are considering i.e, ($ 10^{9}\leq T \leq 10^{12}\, GeV$) and with an absolute neutrino mass $m$ greater than $0.07 eV$. Therefore, in future if the experiments measuring the sum of the neutrino masses $\sum m_{\nu}$ restricts it to the present value of the bound then the degenerate model will also not survive if CP is an exact symmetry of right handed sector and if leptogenesis takes place within the above mentioned temperature range. But this would not be true if the right handed heavy neutrinos are also quasi-degenerate.

\section{Acknowledgement}
We would like to thank Werner Rodejohann for going through the manuscript and 
useful comments.




\begin{thebibliography}{}

\bibitem{Ade:2013zuv} 
  P.~A.~R.~Ade {\it et al.}  [Planck Collaboration],
  arXiv:1303.5076 [astro-ph.CO].

\bibitem{Fukugita:1986hr} 
  M.~Fukugita and T.~Yanagida,
  Phys.\ Lett.\ B {\bf 174}, 45 (1986).

\bibitem{Kuzmin:1985mm} 
  V.~A.~Kuzmin, V.~A.~Rubakov and M.~E.~Shaposhnikov,
  Phys.\ Lett.\ B {\bf 155}, 36 (1985).

 \bibitem{An:2012eh} 
  F.~P.~An {\it et al.}  [DAYA-BAY Collaboration],
  Phys.\ Rev.\ Lett.\  {\bf 108}, 171803 (2012)
  [arXiv:1203.1669 [hep-ex]].

\bibitem{Ahn:2012nd} 
  J.~K.~Ahn {\it et al.}  [RENO Collaboration],
  Phys.\ Rev.\ Lett.\  {\bf 108}, 191802 (2012)
  [arXiv:1204.0626 [hep-ex]].

\bibitem{Feldman:2012qt} 
  G.~J.~Feldman, J.~Hartnell and T.~Kobayashi,
  Advances in High Energy Physics {\bf 2013}, , 475749 (2013)
  [arXiv:1210.1778 [hep-ex]].

\bibitem{Rodejohann:2012xd} 
  W.~Rodejohann,
  J.\ Phys.\ G {\bf 39}, 124008 (2012)
  [arXiv:1206.2560 [hep-ph]].



\bibitem{Pascoli:2006ie} 
  S.~Pascoli, S.~T.~Petcov and A.~Riotto,
  Phys.\ Rev.\ D {\bf 75}, 083511 (2007)
  [hep-ph/0609125].

\bibitem{Branco:2006ce} 
  G.~C.~Branco, R.~Gonzalez Felipe and F.~R.~Joaquim,
  Phys.\ Lett.\ B {\bf 645}, 432 (2007)
  [hep-ph/0609297].

\bibitem{Pascoli:2006ci} 
  S.~Pascoli, S.~T.~Petcov and A.~Riotto,
  Nucl.\ Phys.\ B {\bf 774}, 1 (2007)
  [hep-ph/0611338].

\bibitem{GonzalezGarcia:2012sz} 
  M.~C.~Gonzalez-Garcia, M.~Maltoni, J.~Salvado and T.~Schwetz,
  JHEP {\bf 1212}, 123 (2012)
  [arXiv:1209.3023 [hep-ph]].

\bibitem{Casas:2001sr} 
  J.~A.~Casas and A.~Ibarra,
  Nucl.\ Phys.\ B {\bf 618}, 171 (2001)
  [hep-ph/0103065].

\bibitem{Abada:2006ea} 
  A.~Abada, S.~Davidson, A.~Ibarra, F.~-X.~Josse-Michaux, M.~Losada and A.~Riotto,
  JHEP {\bf 0609}, 010 (2006)
  [hep-ph/0605281].

\bibitem{Petcov:2005jh} 
  S.~T.~Petcov, W.~Rodejohann, T.~Shindou and Y.~Takanishi,
  Nucl.\ Phys.\ B {\bf 739}, 208 (2006)
  [hep-ph/0510404].

\bibitem{Schonert:2005zn} 
  S.~.Schonert {\it et al.}  [GERDA Collaboration],
  Nucl.\ Phys.\ Proc.\ Suppl.\  {\bf 145}, 242 (2005).

\end{thebibliography}
\end{document}